\documentclass[letterpaper,12pt
]{article}   
\usepackage{osajnl2} 

\def\d#1{#1^\dagger}

\begin{document}

\title{Enhanced optical communication and broadband sub-shot-noise interferometry with a stable free-running periodically-poled-$\mathbf{KTiOPO_4}$ squeezer}
\author{Daruo Xie,$^1$ Matthew Pysher,$^1$ Jietai Jing,$^2$ and Olivier Pfister$^{1*}$}
\address{$^1$ Department of Physics, University of Virginia, 382 McCormick Road, Charlottesville, VA 22904-4714, USA}
\address{$^2$ Department of Physics, University of Maryland, College Park, MD 20742-4111, USA}
\address{$^*$Corresponding author: opfister@virginia.edu}

\begin{abstract}
An intrinsically stable type-I optical parametric oscillator was built with a periodically poled ${KTiOPO_4}$ (PPKTP) crystal to generate a stable bright, continuous-wave, broadband phase-squeezed beam. A 3.2 dB sensitivity enhancement of optical interferometry was demonstrated on weak electrooptic modulation signals within a 20 MHz squeezing bandwidth. This also realized a channel capacity increase beyond that of coherent optical communication.
\end{abstract}
\maketitle 

Quasi-phase-matched nonlinear materials such as periodically poled ${KTiOPO_4}$ (PPKTP) are particular cases of photonic crystals that are extremely promising for the enhanced generation of nonclassical states of light, such as squeezed and entangled states. Recently, a record amount of narrowband single-mode squeezing (-9 dB) was observed by use of a PPKTP optical parametric amplifier (OPA) \cite{takeno}. In addition, the elegance and power of quasi-phase-matching \cite{bloembergen,fejer,lifshitz} opens up interesting avenues for quantum information \cite{pooser,menicucci}. The more traditional application of squeezed states to ultrasensitive measurements, e.g., here, interferometry \cite{caves81,xiao,grangier,feng}, is still important. Here we demonstrate a quantum enhancement of the broadband detection sensitivity to optical phase modulation, by use of a PPKTP-based seeded OPA. The intrinsic stability of our OPA makes it possible to make a broadband spectral measurement with squeezed light from the free-running emitter, as opposed to the usual narrowband, phase-scanned, squeezing measurements \cite{takeno}. This result can also be viewed as the realization of squeezed-state detection enhancement of a classical optical communication channel.

\begin{figure*}[htb]
\begin{center}
\includegraphics[width=5.0in]{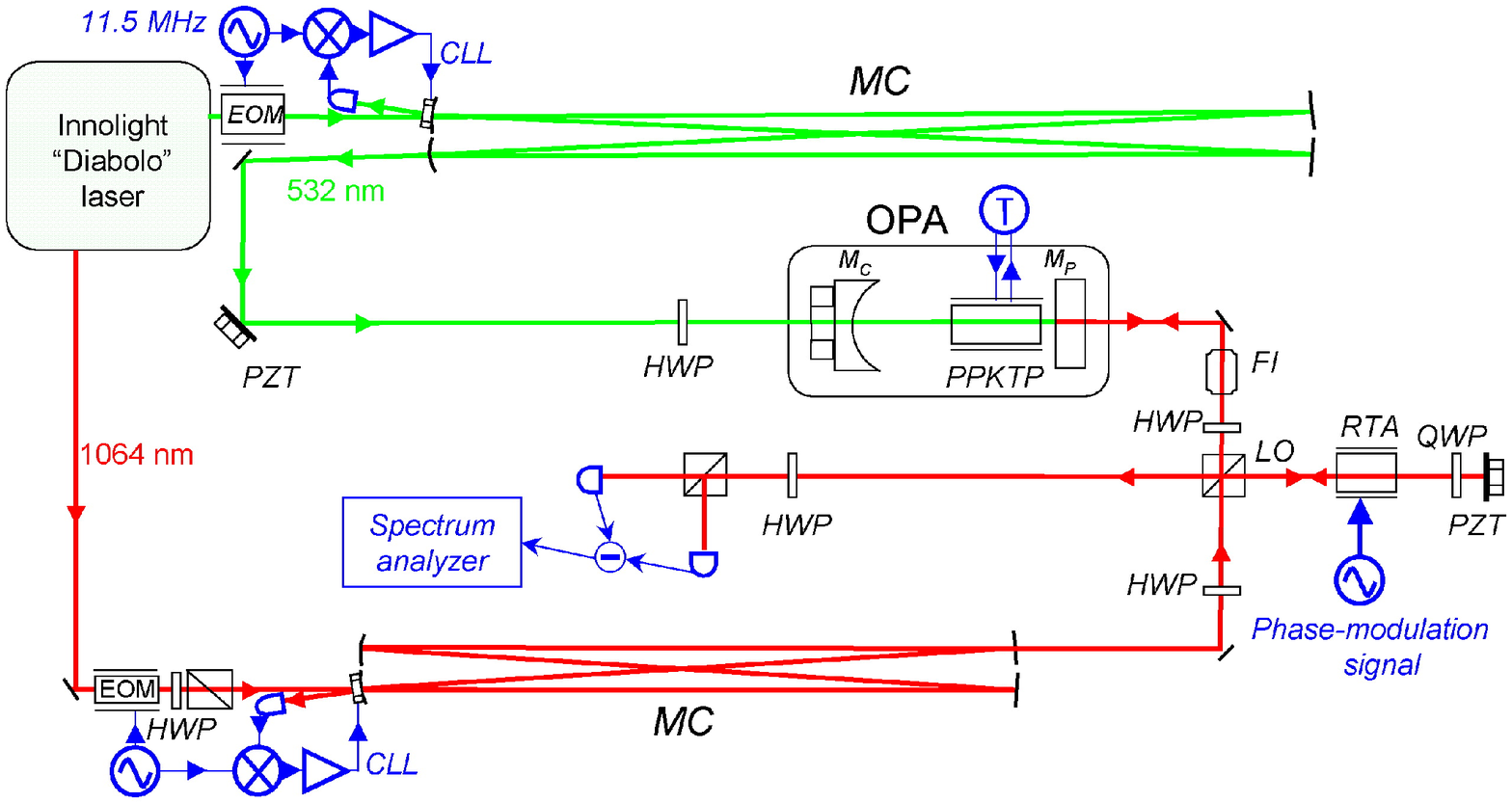}
\end{center}
\caption{Experimental setup. CLL: cavity lock loop, EOM: electrooptic modulator, FR: Faraday rotator ($45^o$), H(Q)WP: half(quarter)-wave plate, LO: local oscillator, MC: mode cleaner, PBS: polarizing beam splitter, PZT: piezoelectric transducer, T: temperature controller.}
\label{setup}
\end{figure*}
The experimental set up is shown schematically in Fig.\ref{setup}. The OPA pump and seed and the local oscillator (LO) beams, at 532 nm, 1064 nm and 1064 nm respectively, were provided by a commercial (Innolight ``Diabolo") neodymium-doped yttrium aluminum garnet (Nd:YAG) monolithic isolated single-mode end-pumped ring (MISER) laser \cite{kane} including an external resonant frequency doubler. All beams were spatially and temporally filtered by two long mode cleaner cavities of 150 kHz full-width at half maximum (FWHM).  The degenerate OPA resonator consisted of a 5mm long KTP crystal, poled with a period of 9 $\mu$m by Raicol, Inc., and of two cavity mirrors, one planar, one spherical, separated by a distance of 52 mm, close to the stability limit. The mirrors were mounted on a super-invar structure (one on a piezoelectric transducer --- PZT), guaranteeing an intrinsically stable resonator. When used as a resonant frequency doubler, the OPO can stay free-running on a single mode for about a minute. The crystal surfaces were coated for high transmission ($R<0.02\%$) of 1064 nm light by MLD Technologies. The PPKTP crystal was mounted in an oven whose temperature is controlled by a temperature lock loop (from Wavelength Electronics) that kept the crystal's operation temperature at $31.10^{\mathrm{o}}$C. This operation temperature corresponds to a phase-matching requirement for this periodically poled crystal to reach its maximum nonlinear efficiency. We measured the crystal's second-harmonic generation efficiency to be $\mathrm{3.83 \times 10^{-3}\ W^{-1}}$. The 532 nm beam was injected through a 50 mm radius of curvature mirror of transmittance $99.4\%$ at 532 nm and $0.008\%$ at 1064 nm. The oscillation threshold pump power for this OPA was $P_{th}=145$ mW. 

The squeezed infrared (IR) wave was extracted through a plane mirror of transmittance $0.06\%$ at 532 nm and $5\%$ at 1064 nm. The PPKTP crystal was  separated from the plane mirror by 0.5 mm.  Scanning the cavity modes with a digital oscilloscope, we measured the OPA cavity's FWHM at 1064 nm to be $21.3(5)$ MHz. Given its free spectral range of $2.66(2)$ GHz, the finesse of the resonator was therefore 125(3), compatible with the value expected from the mirrors' reflectivities alone \{$\mathcal F= \pi(R_1R_2)^{1/4}[1-(R_1R_2)^{1/2}]^{-1}=122$\}. We therefore deduced that the losses in the nonlinear crystal were well below that of the output mirror. The OPA seed and LO IR beams were both provided by the output beam of the 1064 nm mode cleaner, which had a power of $4.2$ mW. A polarizing beam splitter (PBS) was used both to divide this filtered IR beam into the seed and LO beams, with $99.9\%$ given to the LO beam (the seed beam power was 4 $\mu$W) and to recombine the OPA output with the LO, once the latter had been weakly phase-modulated by electro-optic modulation (EOM) in a $\mathrm{RbTiAsO_4}$ (RTA) crystal (made by Coherent Crystal Associates). The subsequent combination of a half-wave-plate and a PBS was used as a balanced beam splitter for balanced homodyne detection and the photocurrent spectrum of the AC interference signal was recorded.

As is standard for a squeezing experiment, control over which quantum field quadrature is squeezed and which one is measured was achieved by respective tuning of the pump and LO optical phases, which were effected by two movable mirrors actuated by piezoelectric transducers (PZTs). The phase relationship $\phi_{\mathit{pump}}-2\phi_{\mathit{seed}}=-\pi/2$ ensured parametric downconversion in the OPA, i.e.\ phase squeezing, while $\theta=\phi_{LO}-\phi_{OPA}=\pm\pi/2$ to measure the phase quadrature and $\phi=0,\pi$ to measure the amplitude quadrature. Both settings were simultaneously verified on DC oscilloscope signals and could be held constant for about a minute once set by hand. Note that all these parameters can be efficiently servo stabilized \cite{schneider,vahlbruch} but one of the goals of this work was to investigate the performance of more rugged systems with high intrinsic stability and a reduced need for active control.

To make noise measurements of phase-squeezed light, we set the 532 nm pump power to $P_{p}=130$ mW, i.e.\ $P_p/P_{th}=0.90$. The OPA output was a squeezed light beam of power $P_{OPA} = 0.16$ mW.  The LO beam had power $P_{LO}=4.2$ mW, i.e.\ $P_{LO} \gg P_{OPA}$. Noise measurements of the squeezed OPA output were then performed by homodyne detection with the LO beam. It is important to point out the detrimental effect that the average power of the OPA beam has on such measurements. If the respective linearized annihilation operators of the OPA and LO modes, phase shifted by $\theta $ are $a=\langle a\rangle+\delta a\equiv\alpha+\delta a$ and $b e^{i\theta} = (\langle b\rangle+\delta b) e^{i\theta} \equiv (\beta+\delta b) e^{i\theta}$, where $\alpha$ and $\beta$ are real numbers, $\langle\delta a\rangle\ll\alpha$, and $\langle\delta b\rangle\ll\beta$, then the balanced homodyne signal gives a measurement of the operator
\begin{eqnarray}
N_- &=& \d{(a+be^{i\theta})}(a+be^{i\theta})/2-\d{(a- b e^{i\theta})}(a-b e^{i\theta})/2  \nonumber\\
&=& \d a b e^{i\theta} + a\d b e^{-i\theta} \nonumber\\
&\simeq& 2\alpha\beta\cos\theta + \alpha\delta B_{-\theta} + \beta\delta A_{\theta},\label{noise}
\end{eqnarray}
where the general quadrature is $A_{\theta}=a e^{-i\theta} + \d a e^{i\theta}$ (same for $b$). The first right-hand term of Eq.\ (\ref{noise}) is the classical interference fringe (DC signal), the second term is an additional shot noise offset, and the third term is the (anti)squeezed noise. Indeed, 
\begin{eqnarray}
V(N_-) = (\Delta N_-)^2 & =&  \alpha^2 \langle\delta B_{-\theta}^2\rangle + \beta^2 \langle\delta A_{\theta}^2\rangle\\
&=& \alpha^2 + \beta^2 (e^{-2r}\sin^2\theta+e^{2r}\cos^2\theta),
\end{eqnarray}
where $r$ is the squeezing parameter. The shot noise measurement is made by blocking the OPA beam. In a phase quadrature measurement ($\theta=\pi/2$), the noise reduction of the OPA light relative to the shot noise is therefore 
\begin{equation}
\frac{V(N_-)_{sq}}{V(N_-)_{sn}} = \frac{\alpha^2 + \beta^2 e^{-2r}}{\beta^2 } 
= \frac{P_{OPA}}{P_{LO}} + e^{-2r}.
\label{corr}
\end{equation}
It is well known \cite{walls} that the OPA squeezing spectrum has the shape of a negative Lorentzian, with a half-width equal to the resonator's FWHM, centered at zero hertz. 
%
\begin{figure}[htb]
\begin{center}
\includegraphics[width=3.0in] {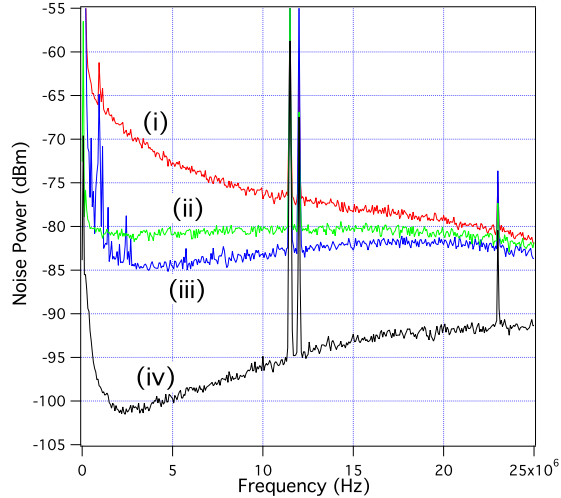}
\end{center}
\caption{Photocurrent power spectrum of the balanced homodyne detection. (i) OPA amplitude quadrature. (ii) Shot noise level. (iii) OPA phase quadrature. (iv) Electronic detection noise (no light). Resolution bandwidth: 30 kHz. Video bandwidth: 10 kHz. 100-sweep average. Peaks at 11.5 and 12 MHz are modulation signals for locking the mode-cleaner cavities.}
\label{sq}
\end{figure}
Figure \ref{sq} shows noise measurement results of the phase and amplitude quadratures and the shot noise (blocked OPA beam). Broadband squeezing is achieved in the detection frequency range from 3 MHz to about 20 MHz. The OPA and detection systems were stable enough for the spectra to be averaged. Since $P_{OPA}/P_{LO}=0.038$ and according to Eq.\ (\ref{corr}), the -3.75 dB average squeezing observed at 3.2 MHz actually corresponds to -4.16 dB. Note that additional optical losses are not taken into account in this calculation. These include the 95\% quantum efficiency of the photodiodes (JDSU ETX-500T), the 96\% homodyne contrast, and the  94\% propagation efficiency from the OPA to the detectors.

We now turn to an application of this source, the improvement of broadband interferometry sensitivity beyond the standard quantum limit \cite{caves81}. Here a differential AC phase shift $\delta\theta(t)=\delta\theta_0\cos\Omega t$ was written between the LO and OPA beams by way of an RTA EOM. This phase shift was then interferometrically detected by the homodyne beat of the two beams (Fig.~\ref{setup}). Two situations were implemented: {\em (i)} Inactive OPA (pump beam blocked) with the OPA beam being the mere seed beam reflected from the off-resonant OPA cavity. {\em (ii)} Active OPA with the OPA beam phase-squeezed as before. Note that a bright squeezed beam is required for such an experiment, which is different from previous interferometry experiments with squeezed vacuum \cite{xiao,grangier}. 

In a RTA crystal (symmetry group $mm2$), EO modulation along the $Z$ principal axis yields \cite{yariv}
\begin{equation}
\delta\theta_0=(n_{Z}^{3}r_{33}E_{Z}+n_{Y}^{3}r_{23}E_{Z})\ \frac{l\pi}{\lambda}
\end{equation}
where $n_{Z}, n_{Y}$ are the indices of refraction of RTA, $r_{33}=36.7$ pm/V and $r_{23}=15.7$ pm/V are electrooptic coefficients of the RTA crystal, and $E_{Z}$ is the electric field generated by an AC signal generator. In the experiment, the unmodulated OPA beam was Z-polarized while the LO beam was Z-polarized in the first pass through the EOM and Y-polarized in the second pass.

\begin{figure}[htb]
\begin{center}
\includegraphics[width=3.0in] {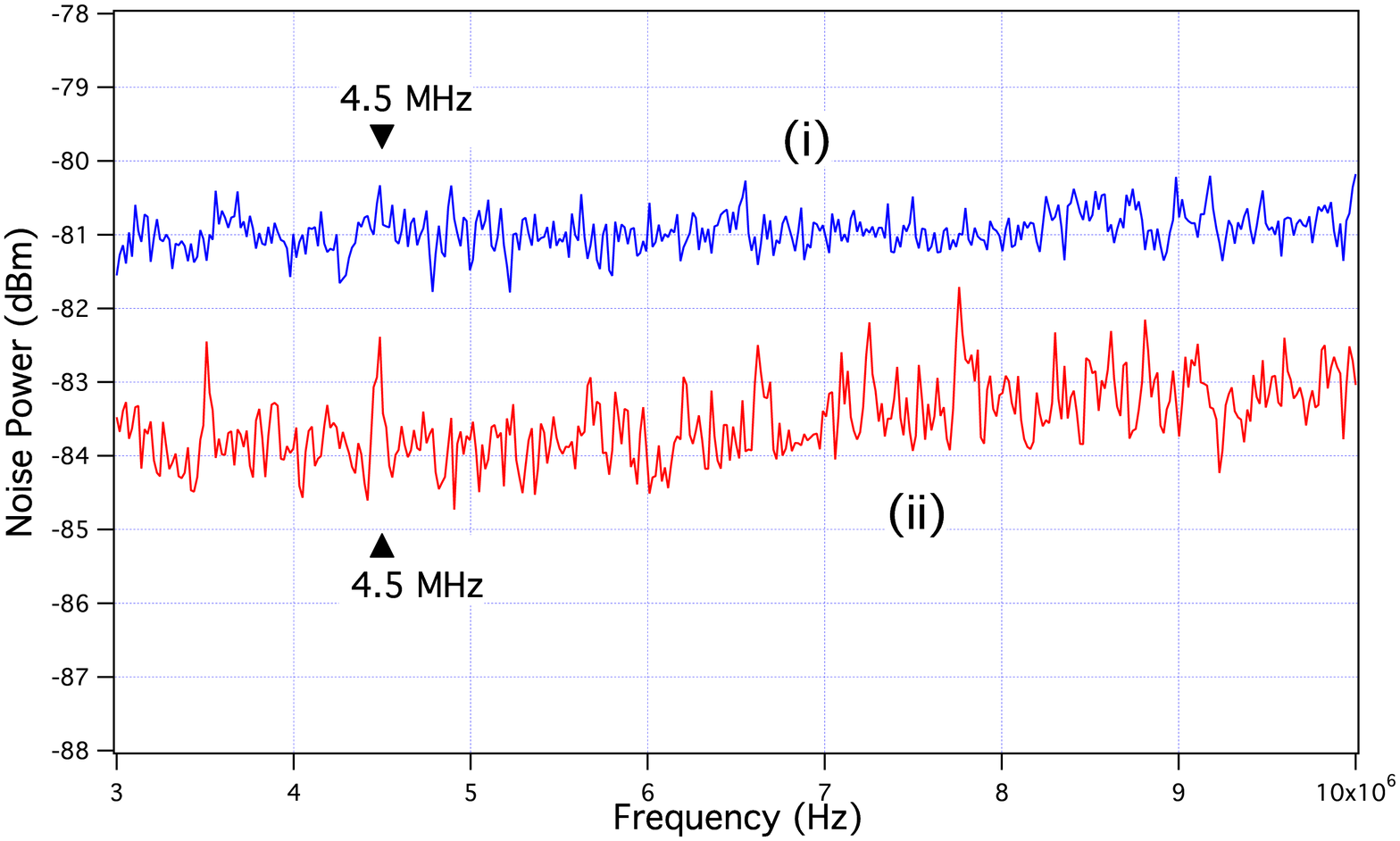}
\end{center}
\caption{Photodetection noise with a small phase modulation at $\Omega/2\pi=4.5$ MHz. {\em (i)} LO beam interferes with a coherent beam. {\em (ii)} LO beam interferes with a squeezed beam. In both cases, the average optical power $P_{OPA}\propto\alpha^2$ is the same. Resolution bandwidth: 30 kHz. Video bandwidth: 10 kHz.  100-sweep average.}
\label{enh}
\end{figure}

Figure \ref{enh} shows the homodyne measurement results, still in free-running conditions with only the crystal temperature stabilized in the OPO. In situation {\em (i)}, we took care to match the power of the reflected seed beam to that of the OPA beam (0.16 mW) in situation {\em (ii)}, so as to have an accurate comparison of experimental conditions and identical shot noise levels. The difference between the two situations was, of course, the presence of vacuum fluctuations in {\em (i)} and of phase squeezing in {\em (ii)}. As can be clearly seen in Fig.\ \ref{enh}, the minute AC modulation peak (which can also be seen as an AC quantum noise variation) is invisible in classical interferometry  {\em (i)} but clearly revealed by phase squeezing  {\em (ii)}. Note that the observed squeezing is only -3 dB here, probably due to small drifts in the unstabilized phases \cite{takeno}. We now turn to the power offset correction. To the difference of the initial squeezing measurement, the shot noise trace in Fig.~\ref{enh} is obtained without blocking the OPA beam, but by substituting to it a classical beam of equal average power (blocked OPA pump). Equation (\ref{corr}) therefore becomes
\begin{equation}
\frac{V(N_-)_{sq}}{V(N_-)_{sn}} = \frac{\alpha^2 + \beta^2 e^{-2r}}{\alpha^2 + \beta^2 }, 
\end{equation}
which yields a corrected squeezing value of -3.2 dB. Moreover, this broadband measurement gives access to a squeezing bandwidth of about 20 MHz, which can be used for detecting more complicated modulation spectra. 

This result is also a demonstration of classical channel capacity enhancement in coherent optical communication. Indeed, this broadband modulation of the coherent LO beam is simply AM or FM encoding. When the encoded signal is weak and buried in the shot noise, we can see that homodyne detection with a coherent bright squeezed beam can bring the signal out. The only requirement for successful decoding is the {\em classical} coherence of the squeezed and modulated beams ($\theta=\pi/2$ well defined, i.e.\ not randomly fluctuating). Because of this, our scheme does not allow secure communication: an eavesdropper can simply pick off part of the encoded LO and phase-lock it to their OPA bright seed beam. Then their OPA squeezed beam will be ready to be used for detection.

It is instructive to put this scheme in perspective with quantum information ones: even though a nonclassical state of light is used here to enhance sensitivity, and hence channel capacity, it does not offer the security of dense coding \cite{li} or quantum correlation \cite{pereira} because these are based on shared quantum correlations between sender and receiver. One cannot successfully eavesdrop on quantum correlations since picking off a signal beam all but contaminates its quantum statistics with vacuum fluctuations.

\begin{figure}[htb]
\begin{center}
\includegraphics[width=3.0in] {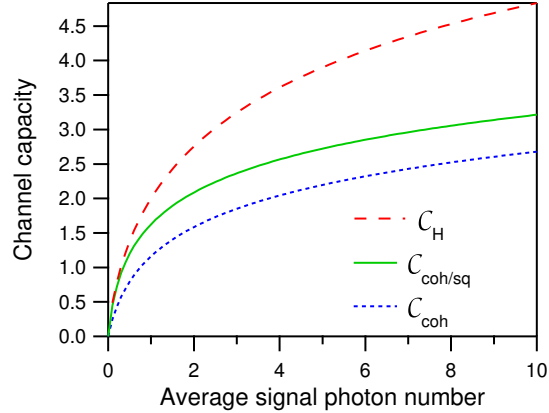}
\end{center}
\caption{Channel capacity versus average photon number in different cases. Dotted line: classical coherent limit. Solid line: -3.2 dB squeezed limit (this work). Dashed line: Holevo bound.}
\label{cc}
\end{figure}

Our scheme nevertheless provides a significant increase of the channel capacity over coherent states. The squeezed and classical channel capacities of Gaussian Wigner functions are given by
\cite{ralph}
\begin{equation}
\mathcal C = \frac 12\log_2\left(1+\frac{V_S}{V_N}\right),
\end{equation}
where $V_S$ and $V_N$ are, respectively, the signal and noise variances of the quantum field amplitude. For a coherent state, $V_N=1$. In our case, the squeezed noise yields $V_N=\exp(-2r)=0.47$ for $-3.2$ dB squeezing. One can then rewrite $\mathcal C$ in terms of the spectral average photon number in the signal beam \cite{webb}. 
\begin{equation}
\bar n=\langle \d aa\rangle=\frac 14 (V_0+V_{\pi/2}) - \frac 12,
\end{equation}
where $V_{\theta}=\langle|\delta A_{\theta}|^2\rangle$. The LO beam has $V_0=V_S+1$, $V_{\pi/2}=1$, hence $\bar n=V_S/4$ and 
\begin{equation}
\mathcal C_{\mathit{coh}} = \frac 12\log_2\left(1+4\bar n\right). 
\end{equation}
In our case, however, 
\begin{equation}
\mathcal C_{\mathit{coh/sq}} = \frac 12\log_2\left(1+4 e^{2r} \bar n\right). 
\end{equation}
One can also compare with the case of a signal-encoded squeezed beam \cite{ralph}, for which $V_0=V_S +\exp(-2r)$, $V_{\pi/2}=\exp(2r)$, which gives $\bar n= V_S/4 +\sinh^2r$ and \begin{equation}
\mathcal C_{\mathit{sq}} = \frac 12\log_2\left[1+4 e^{2r}(\bar n-\sinh^2r)\right],
\end{equation}
where $\bar n>\sinh^2r$ is ensured by the presence of an encoded signal ($V_S\neq 0$). Figure \ref{cc} displays the channel capacities versus signal strength for our experiment. Note that neither can exceed the Holevo bound,
\begin{equation}
\mathcal C_{\mathit{H}} = (1+\bar n)\log_2(1+\bar n) - \bar n\log_2\bar n,
\end{equation}
which can only be breached by use of quantum entanglement, i.e.\ dense coding \cite{braunstein}.

In conclusion, we have generated a bright, broadband, squeezed light beam, by use of an intrinsically stable, free-running PPKTP degenerate OPA, and applied it to broadband interferometry 3.2 dB below the shot noise. Note that the squeezing bandwidth can be considerably increased very simply by using a shorter OPA cavity, e.g.\ a monolithic crystal resonator. This finds a direct application as classical channel capacity enhancement in broadband coherent optical communication. This work was supported by NSF grant Nos.\ PHY-0555522 and CCF-0622100.


\begin{thebibliography}{99}
\bibitem{takeno} Y. Takeno, M. Yukawa, H. Yonezawa, and A. Furusawa, ArXiv e-print quant-ph/0702139.
\bibitem{bloembergen} J.A. Armstrong, N. Bloembergen, J. Ducuing, and P.S. Pershan, Phys.\ Rev.\ {\bf 127}, 1918 (1962).
\bibitem{fejer} M.M. Fejer, G.A. Magel, D.H. Jundt, and R.L. Byer, \jqe {\bf 28}, 2631 (1992).
\bibitem{lifshitz} R. Lifshitz, A. Arie, and A. Bahabad, \prl {\bf 95}, 133901 (2005).
\bibitem{pooser} R.C. Pooser and O. Pfister, \ol {\bf 30}, 2635 (2005).
\bibitem{menicucci} N.C. Menicucci, S.T. Flammia, H. Zaidi, and O. Pfister, ArXiv preprint quant-ph/0703096.
\bibitem{caves81} C.M. Caves, \prd {\bf 23}, 1693 (1981).
\bibitem{xiao} M. Xiao, L.-A. Wu, and H.J. Kimble, \prl {\bf 59}, 278 (1987).
\bibitem{grangier} P. Grangier, R.E. Slusher, B. Yurke, and A. LaPorta, \prl {\bf 59}, 2153 (1987).
\bibitem{feng} S. Feng and O. Pfister, \ol {\bf 29}, 2800 (2004).
\bibitem{kane} T.J. Kane and R.L. Byer, \ol {\bf 10}, 65 (1985).
\bibitem{schneider} K. Schneider, R. Bruckmeier, H. Hansen, S. Schiller, and J. Mlynek, \ol {\bf 21}, 1396 (1996).
\bibitem{vahlbruch} H. Vahlbruch, S. Chelkowski, B. Hage, A. Franzen, K. Danzmann, and R. Schnabel, \prl {\bf 97}, 011101 (2006).
\bibitem{walls} D.F. Walls and G.J. Milburn, {\em Quantum Optics}, Springer (1994).
\bibitem{yariv} A. Yariv, {\em Quantum electronics}, 3rd Ed., Wiley (1988).
\bibitem{pereira} S.F. Pereira, Z.Y. Ou, and H.J. Kimble, \pra {\bf 62}, 042311 (2000).
\bibitem{li} X. Li, Q. Pan, J. Jing, J. Zhang, C. Xie, and K. Peng, \prl {\bf 88}, 047904 (2002).
\bibitem{ralph} T.C. Ralph and E.H. Huntington, \pra {\bf 66}, 042321 (2002).
\bibitem{webb} J.G. Webb, T.C. Ralph, and E.H. Huntington, \pra {\bf 73}, 033808 (2006).
\bibitem{braunstein} S.L. Braunstein and H.J. Kimble, \pra {\bf 61}, 042302 (2000).
\end{thebibliography}
\end{document}